\documentclass[aps,prl,twocolumn,groupedaddress,showpacs]{revtex4}

\usepackage{psfig,graphics,graphicx}

\begin{document}

\title{Defect-mediated turbulence in systems with local deterministic chaos}

\affiliation{}
\author{J\"orn Davidsen}
\email[]{jdavidse@chem.utoronto.ca}
\author{Raymond Kapral}
\email[]{rkapral@chem.utoronto.ca}
\affiliation{Chemical Physics
Theory Group, Department of Chemistry, University of Toronto,
Toronto, ON M5S 3H6 Canada}

\date{\today}

\begin{abstract}
We show that defect-mediated turbulence can exist in media where
the underlying local dynamics is deterministically chaotic. While
many of the characteristics of defect-mediated turbulence, such as
the exponential decay of correlations and a squared Poissonian
distribution for the number of defects, are identical to those
seen in oscillatory media, the fluctuations in the number of
defects differ significantly. The power spectra suggest the
existence of underlying correlations that lead to a different and
non-universal scaling structure in chaotic media.
\end{abstract}

\pacs{47.54.+r, 82.40.Ck}

\maketitle

Weakly driven, dissipative, pattern-forming systems often exhibit
the spatiotemporal chaos in the form of defect-mediated
turbulence, where the dynamics of a pattern is dominated by the
perpetual motion, nucleation and annihilation of point defects (or
dislocations) \cite{coullet89}. Examples can be found in striped
patterns in wind driven sand, electroconvection in liquid crystals
\cite{rehberg89}, nonlinear optics \cite{ramazza92}, fluid
convection \cite{morris93,daniels02}, autocatalytic chemical
reactions \cite{ouyang96} and Langmuir circulation in the oceans
\cite{haeusser97}. These results suggest that the dynamics of
these very different systems can be characterized by a universal
description which is based only on the defect dynamics.

A statistical description of defect
mediated turbulence based on a simple model for the defect
dynamics was given by Gil, \emph{et al.} \cite{gil90}. Treating
the defect pairs as statistically independent entities, the
nucleation rate for pairs of defects was taken to be independent
of the number of pairs $n$ and, based on the topological nature of
the defects, the annihilation rate was taken to be proportional to
$n^2$. This directly led to a squared Poissonian distribution for
the probability distribution function (PDF) of $n$. Gil, \emph{et
al.} found that simulations of a spatiotemporal chaotic state of
the complex Ginzburg-Landau equation (CGLE), which is the
prototype of oscillatory media, agreed with their prediction.
Rehberg, \emph{et al.} \cite{rehberg89} measured the PDF of $n$
for defect-mediated turbulence in electroconvection of nematic
liquid crystals and found it to be consistent with the predicted
squared Poissonian distribution. Later, Ramazza, \emph{et al.}
\cite{ramazza92} investigated a defect turbulent state in optical
patterns and found that their data were not conclusive. Very
recently, Daniels and Bodenschatz \cite{daniels02} studied the
defect-mediated turbulent state of undulation chaos in inclined
layer convection of a fluid and found that the observed pair
nucleation and annihilation rates agree with the theoretical
predictions in \cite{gil90}. They also derived the PDF of $n$ for
the case that single defects can leave and enter through the
boundaries and found agreement with their measurements. When
boundary effects are negligible, this PDF reduces to the squared
Poissonian distribution.

The experimental and theoretical studies of defect-mediated
turbulence have focused exclusively on first order statistics like
the PDF of $n$ and creation and annihilation rates. Moreover,
theoretical investigations have been carried out only for media
with underlying oscillatory dynamics, generically described by the
CGLE. Yet, in many cases the local dynamics may differ from simple
oscillatory behavior; instead, complex-periodic or even chaotic
attractors may exist (see, e.g., Ref. \cite{goryachev00}).
Chemically reacting systems, notably the Belousov-Zhabotinsky (BZ)
reaction, are known to exhibit deterministic chaos \cite{scott}.
This leads to a variety of new spatiotemporal states
\cite{park99}. In this Letter, we show that defect-mediated
turbulence can exist in media where the underlying local dynamics
is chaotic and that second-order statistics can be used to
distinguish between different media. This implies that a universal
description of defect-mediated turbulence cannot encompass
second-order or higher correlations.

\begin{figure*}[htbp]
 \includegraphics[width=\textwidth]{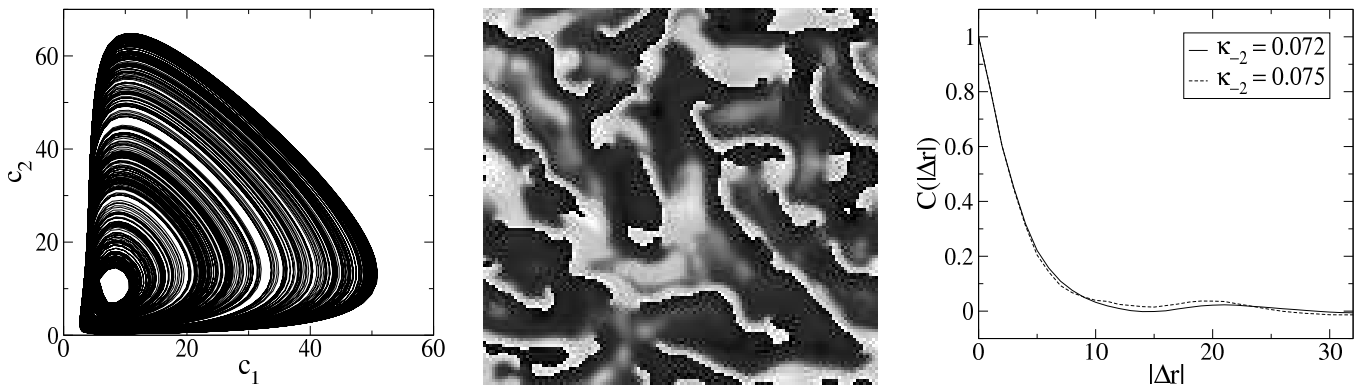}%
 \caption{\label{all} Left: Projection of the chaotic attractor
 in the $(c_1,c_2)$-plane of the homogeneous
 WR model for $\kappa_1=31.2,
 \kappa_{-1}=0.2, \kappa_2=1.45, \kappa_{-2}=0.072, \kappa_3=10.8,
 \kappa_{-3}=0.12, \kappa_4=1.02, \kappa_{-4}=0.01, \kappa_5=16.5,
 \kappa_{-5}=0.5$.
 Snapshots of the inhomogeneous ${\bf c}$ field in the defect
 turbulent state closely resemble this attractor.
 Center: Phase field
 for $L = 128$ \cite{numerics}.
 Right:
 Correlation function in the defect-mediated turbulent state
 for $L = 128$ and different values of $\kappa_{-2}$. Note that
 $\bar n = 15.84$ for $\kappa_{-2} = 0.072$ and $\bar n = 8.81$ for
 $\kappa_{-2} = 0.075$.}
 \end{figure*}

Our focus is on systems where the dynamics of the spatially
homogeneous system, described by ordinary differential equations,
has a deterministic chaotic attractor; hence, at least three phase
space variables are required in contrast to the two-variable
descriptions of simple oscillatory media. A specific example of
such a chaotic system is the Willamowski-R\"ossler (WR)
reaction-diffusion model \cite{willamowski80} given by
\begin{eqnarray}
\partial_t {\bf c} ({\bf r},t) &=&  {\bf R} [{\bf c} ({\bf r},t)] +
D \nabla^2 {\bf c} ({\bf r},t),
\end{eqnarray}
where $R_1 = \kappa_1 c_1 - \kappa_{-1} c_1^2 - \kappa_2 c_1 c_2 +
\kappa_{-2} c_2^2 - \kappa_4 c_1 c_3 + \kappa_{-4}$, $R_2 =
\kappa_2 c_1 c_2 - \kappa_{-2} c_2^2 - \kappa_3 c_2 + \kappa_{-3}$
and $R_3 = - \kappa_4 c_1 c_3 + \kappa_{-4} + \kappa_5 c_3 -
\kappa_{-5} c_3^2$. These rate equations were derived from the
mass action kinetics of a reaction scheme with quadratic kinetics
where certain pool species are taken to be fixed. Here $c_i({\bf
r},t)$ is the local concentration of species $i$ at site ${\bf r}$
in a two-dimensional space of size $L^2$ with periodic boundary
conditions. The parameters $\kappa_{\pm j}$ are rate coefficients
that contain the concentrations of the pool species that are fixed
to maintain the system out of equilibrium. The diffusion
coefficients $D$ of all three species are taken to be equal. Even
though the WR model is very simple, it exhibits a
phenomenology~\cite{goryachev96} with many features in common with
those observed in chemical experiments on the BZ
reaction~\cite{park99}. Consequently, this model may be expected
to capture the qualitative features of chemical systems whose
chaotic attractors arise from a period-doubling cascade. The
chaotic attractor for a certain set of rate coefficients is shown
in the left panel of Fig.~\ref{all}.

A defect is characterized by its integer topological charge
$m_{top}$ which is defined by \cite{mermin79}
\begin{equation}
\frac{1}{2\pi }\oint \nabla \phi ({\bf r},t)\cdot d{\bf l}=\pm
m_{top},
\end{equation}
where $\phi ({\bf r},t)$ is the local phase and the integral is
taken along a closed curve surrounding the defect. A topological
defect corresponds to a point in the medium where the local
amplitude is zero and the phase is not defined. Typically only
topological defects with $m_{top} = \pm 1$ are observed. One-armed
spiral waves with such defects at their centers are the only
stable spiral waves for the CGLE \cite{hagan82}. The phase has to
be defined in order to apply the notion of defects to chaotic
media. This is not a trivial issue and for many chaotic systems it
is impossible to introduce a phase field. The rather simple shape
of the WR chaotic attractor (see left panel of Fig.~\ref{all})
which arises from a period-doubling cascade admits a simple
definition of the phase and the WR model belongs to the class of
chaotic-oscillatory media \cite{pikovsky}. We chose $\phi ({\bf
r},t) = \arctan [(c_2({\bf r},t) - c_2^0) / (c_1({\bf r},t) -
c_1^0)]$ with $(c_1^0, c_2^0) = (8.0, 9.0)$ as the center of
rotation. For such simple chaotic attractors, the particular
choice of a phase variable for chaotic systems does not influence
the results \cite{josic01}. In the center panel of Fig. \ref{all},
the phase field of the WR medium is shown for a certain set of
parameters \cite{numerics}. The topological defects can be
identified as the termini of the white equiphase contour lines. As
for the CGLE in the defect-mediated turbulent state
\cite{chate96,aranson02}, the defects in the WR system for these
parameter values rarely emit waves. They behave as passive objects
and are merely advected by the surrounding chaotic fluctuations.
The right panel in Fig.~\ref{all} shows that the correlation
function $C(|{\bf \Delta r}|) = \langle Re[A({\bf r},t) \bar
A({\bf r + \Delta r},t)] \rangle_{{\bf r},t}$, where $A({\bf
r},t)$ is the complex amplitude in the $(c_1,c_2)$-plane with
respect to the center of rotation, decays exponentially with a
very short characteristic length scale, verifying the existence of
a turbulent state. This is the most typical characteristic for
defect-mediated turbulence and has been found in the CGLE as well
\cite{coullet89}.

 \begin{figure}[htbp]
 \vspace*{1.3cm}
 \psfig{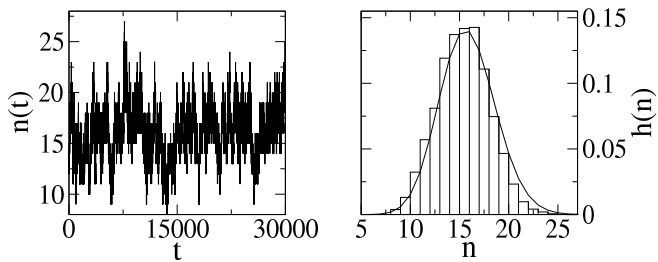}
 \caption{\label{timeseries}Left: Time series of the number of
 defect pairs $n$ for the parameters given
 in Fig. \ref{all} and $L = 128$. Right: Normalized histogram
 $h(n)$ of $n$. The solid curve is the corresponding squared
 Poissonian  distribution.}
 \end{figure}

\begin{figure}[htbp]
 \vspace*{1.3cm}
 \psfig{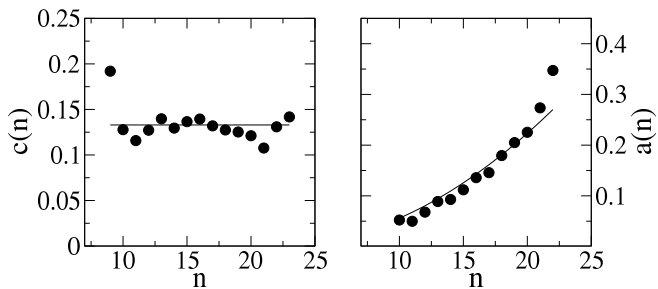}
 \caption{\label{rates}Left: $c(n)$ in the WR system for the
 parameters given in Fig. \ref{all}
 and $L = 128$. The solid line corresponds to a constant creation
 rate.
 Right: $a(n)$ in the WR system for the same parameters.
 The solid line corresponds to an increase in the annihilation
 rate proportional to
 $n^2$.}
 \end{figure}

The fluctuations in the number of pairs of topological defects
shown in the left panel of Fig.~\ref{timeseries} provide further
evidence that a defect-mediated turbulent state can exist in
chaotic-oscillatory media. The total number of defects in the
medium is exactly twice the number of pairs because the net
topological charge is conserved and equal to zero due to the periodic
boundary conditions. Hence, topological defects can only be
created and annihilated in pairs of opposite topological charge.
One can easily derive a PDF $p(n)$ for the number of defects
provided that the defects are statistically independent entities
\cite{gil90,daniels02}. In the stationary state and for periodic
boundary conditions, the master equation reduces to $p(n) = p(n-1)
c(n-1)/a(n)$ where $c(n)$ and $a(n)$ are the creation and
annihilation rates, respectively. Provided that $c(n)=c={\rm
const}$ and $a(n)=a n^2$, $p(n) \propto (c/a)^n / (n!)^2$. The
right panel of Fig.~\ref{timeseries} shows that the PDF for the WR
model agrees with the predicted form of a squared Poissonian
distribution reasonably well. Moreover, the assumptions leading to
this distribution seem to be justified: the annihilation rate
scales approximately with $n^2$ and the creation rate is
approximately independent of $n$ as can be deduced from
Fig.~\ref{rates}.

\begin{figure}[htbp]
 \psfig{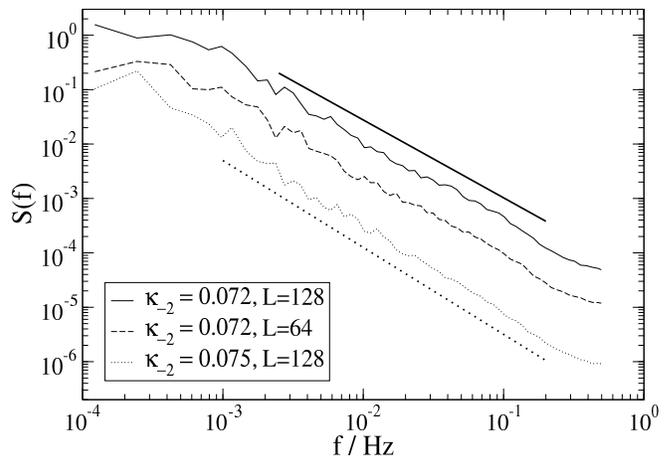}
 \caption{\label{power_spectrum}Power spectrum of the signal $n(t)$
 for different values of $\kappa_{-2}$ and $L$.
 For clarity, the curve for $\kappa_{-2}=0.075$ has been shifted
 down by one decade. The thick lines are to guide the eye.
 The thick solid line
 decays with $\gamma=1.43$ and the thick dotted line with
 $\gamma=1.60$. Note that the average period for one oscillation
 is $\bar{T} = 0.998$
 for $\kappa_{-2}=0.072$ and $\bar{T} = 0.980$ for
 $\kappa_{-2}=0.075$.}
 \end{figure}

The fluctuations in the number of defects $n$ do not have the
properties one would expect if they arose from independent random
events. If this were the case the power spectrum $S_L(f) =
\lim_{T\to\infty} 1/2T \mid \int\limits_{-T}^{T} dt\; n(t)
\exp^{-i 2\pi f t} \mid^2$ for a system of size $L$ would have a
Lorentzian shape. The power spectrum does not have this form for
the defect-mediated turbulence in the WR reaction-diffusion
system. Figure~\ref{power_spectrum} shows that $S_L(f) \propto
1/f^\gamma$ for intermediate frequencies with an exponent $\gamma$
that is far from the value $\gamma=2$ expected for a Lorentzian
shape. For $k_{-2} = 0.072$, we find $\gamma = 1.43$ and, for
$k_{-2} = 0.075$, $\gamma = 1.60$. Although in both cases the
system exhibits defect-mediated turbulence, the exponents are
significantly different from each other. This implies that
different chaotic-oscillatory media can have different
second-order statistics. To confirm that these results are not
specific to the WR system, we have also studied the autocatalator
reaction-diffusion system~\cite{auto,kapral97} which also has a
chaotic attractor arising from a period-doubling cascade but with
very different Lyapunov spectra. We find a power-law decay with
values of $\gamma$ that are similar to those reported here for the
WR model \cite{footnote}. These results suggest that the power
spectrum of $n(t)$ may exhibit power-law decay with non-trivial
exponents for defect-mediated turbulent states when the local
dynamics exhibits deterministic chaos.

For large enough system sizes, $S_L(f) \propto L^2$ for all $f$;
thus, the power spectrum for large systems can be considered to be
the superposition of the power spectra of subsystems. Such
behavior is expected in view of the short correlation length. This
implies that the standard deviation $\Sigma = (\langle
n^2 \rangle - \langle n\rangle^2)^{1/2}$ scales as $\sqrt{\langle
n \rangle}$, which is proportional to $L$. Such a scaling is a
consequence of the law of large numbers and has been observed for
the CGLE as well. It follows that $\gamma$ and the low-frequency
cutoff are independent of $L$. The low-frequency cutoff is due to
the fact that $n(t)$ is bounded. Its location depends on the
density of defects in the medium. For lower densities the cutoff
moves to lower frequencies.

To understand how the non-trivial correlations in $n(t)$ at
intermediate time scales arise, we have analyzed the series of
waiting times between consecutive creation events and consecutive
annihilation events separately. In both cases, the waiting times
are exponentially distributed and statistically uncorrelated as
for a random walk. This implies that the correlations in $n(t)$
are due to the interaction of creation and annihilation events.

\begin{figure}
 \psfig{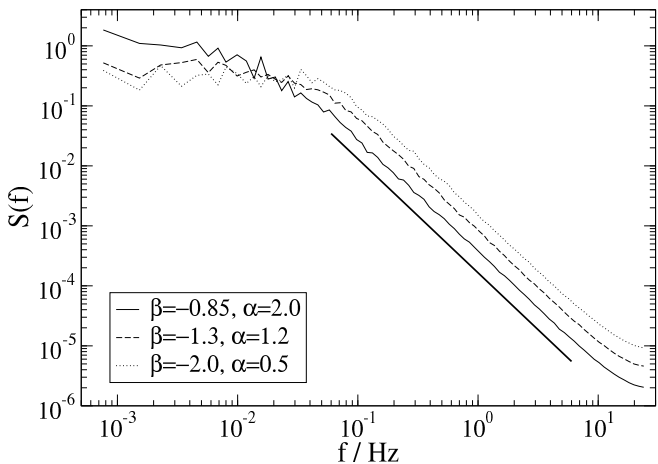}
 \caption{\label{power_spectrum_cgl}Power spectrum of $n(t)$
 for different parameters in the CGLE and $L=128$.
 The thick line is to guide the eye and decays with $\gamma=1.9$.
 Note that $\bar T = 12.7, 8.40, 5.43$ from highest
 to lowest $\alpha$.}
 \end{figure}

To compare our results for the power spectrum of $n(t)$ in chaotic
media with those in oscillatory media, we have simulated the CGLE
\cite{coullet89,aranson02}
\begin{equation}
\partial_t A = A + (1 + i \alpha) \nabla^2 A - (1 + i \beta) |A|^2
A,
\end{equation}
in a domain of size $L^2$ with periodic boundary conditions
\cite{numerics_cgl}. Here, $A$ is the complex amplitude field and
$\alpha$ and $\beta$ are control parameters. The power spectrum of
the CGLE in the defect-mediated turbulent state has a power-law
decay for intermediate frequencies and the scaling of $S_L(f)$
with $L$, as well as the dependence of the low-frequency cutoff on
the density of defects, agree with those in the WR medium. Yet, as
Fig.~\ref{power_spectrum_cgl} shows, the exponent $\gamma = 1.9$
which is very close to that expected for a Lorentzian form if the
dynamics of $n(t)$ is a simple random walk in a bounded domain. It
is very different from the values observed for the chaotic media
we analyzed. Thus, different media can have different second-order
statistics for $n(t)$ depending on the underlying local dynamics.
This argues against a universal description of defect-mediated
turbulence.

We have shown that defect-mediated turbulence can arise both in
oscillatory and chaotic-oscillatory media and, thus, applies to a
broad range of systems. While most common diagnostic
measures do not allow one to distinguish between oscillatory and
various chaotic media, the fluctuations in the number of defects
can be different for different media: for the CGLE they resemble
the form of a bounded random walk; for the WR and autocatalator
models pronounced non-trivial correlations exist. Our results may
be tested experimentally on systems like the Belousov-Zhabotinsky
reaction whose local temporal dynamics can be complex-periodic or
even chaotic \cite{park99}, and where defect-mediated
turbulence has been observed \cite{ouyang96}.

This work was supported in part by a grant from the Natural
Sciences and Engineering Council of Canada.


\begin{thebibliography}{28}
\bibitem{numerics}
    The numerical integrations were performed by a simple forward Euler scheme
    with time step $\Delta t = 0.0002$, grid resolution $\Delta x =
    1$ and a five-point Laplacian.
\bibitem{footnote}
    For the parameters given in Ref. \cite{kapral97} and $\kappa_2 = 11.52$,
    we find $\gamma = 1.40$. The positive Lyapunov
    exponent in the homogeneous system is approximately a factor of two
    smaller than in the WR model for the parameters we have considered here.
    This and the $\gamma$ variations in the WR systems with similar
    Lyapunov exponents suggests that $\gamma$ does not depend solely on the largest
    Lyapunov exponent.
\bibitem{numerics_cgl}
    The numerical integrations were performed by a simple forward Euler scheme
    with time step $\Delta t = 0.02$, grid resolution $\Delta x = 0.5$
    and a nine-point Laplacian.
\end{thebibliography}
\end{document}